\documentstyle[epsfig]{aipproc}

\begin{document}
\noindent hep-ph/9706447 \hfill {Saclay--SPhT/T97--62}

\vskip -20pt
\title{Advanced Techniques for Multiparton Loop Calculations: A Minireview\thanks{Presented by D. A. Kosower at 
the 5${}^{\rm th}$ International Workshop on Deep Inelastic Scattering
and QCD, Chicago, Illinois, April 14--18, 1997}}

\author{Zvi Bern$^a$, Lance Dixon$^b$, David C. Dunbar$^c$,
and David~A.~Kosower$^d$}
\address{$^a$ Department of Physics, UCLA, Los Angeles, CA 90024\\
$^b$SLAC, Stanford, CA 94309 \\
$^c$University College of Swansea, Swansea, UK\\
$^d$Service de Physique Th\'eorique, CEA---Saclay, 
F--91191 Gif-sur-Yvette cedex, 
France\thanks{Laboratory of the {\it Direction des Sciences de la Mati\`ere\/}
of the {\it Commissariat \`a l'Energie Atomique\/} of France.}}

\maketitle

\begin{abstract}
We present an overview of techniques developed in recent years for
the efficient calculation of one-loop multiparton amplitudes, 
in particular those relying on unitarity and collinear factorization.
\end{abstract}

\section*{Introduction}

Much of the experimental effort in high-energy physics today is 
directed at searching for new physics beyond the standard model.
Successful searches will require a detailed understanding of known physics.
At high-energy colliders, particularly hadron-hadron colliders, this
requirement implies above all the need for a detailed understanding 
of perturbative QCD.  A detailed theoretical picture of dijet production,
for example, is necessary if we are to use it in constraining the
gluon distribution.  The search for new physics in the single-jet
inclusive distribution likewise requires detailed theoretical calculations
(in addition to better measurements of the parton distributions than
have been available heretofore).  Other measurements which rely on
our detailed understanding of perturbative QCD include the $W$ mass,
$W$+jet ratios, and top production in semi-leptonic and purely hadronic
channels.

\def\yir{y_{\rm  IR}} 

In the context of jet physics at high-energy colliders, next-to-leading
order calculations represent the first step in providing such a detailed
picture.  Unlike the case of more inclusive measurements, such as
the total hadronic cross section in $e^+e^-$ annihilation, the perturbative
expansion for jet observables contains both ``ultraviolet'' and ``infrared''
logarithms.  The former arise from the truncation of the perturbation
series at a fixed order.  They manifest themselves in the (unphysical)
renormalization-scale
dependence of theoretical predictions.  This dependence is numerically strong
at leading order, because the coupling is still relatively large, and runs
relatively quickly.  Next-to-leading order effects compensate this
sensitivity, and in practice reduce significantly the undesired
renormalization-scale dependence.  The second type of logarithms arise because
of the presence of different scales, a hard scale characterizing the scattering,
and softer scales characterizing the size of a jet.  They modify the perturbative
expansion from one solely in $\alpha_s$ to one in $\alpha_s \ln^2 \yir$ and
$\alpha_s\ln \yir$ as well, where $\yir$ is the ratio of the different scales.
Only at next-to-leading order can we justify quantitatively the 
harmless nature of these logarithms, and thereby the applicability of
perturbation theory.

\section*{Organizing a Calculation}

\def\ellbar{\overline{\ell}}
At leading order, producing numerical predictions for an 
$n$-jet process in lepton-hadron collisions requires knowledge of the parton distribution
functions of the proton; of $\alpha_s$; and of the tree-level matrix
element for the $\ell + x\rightarrow \ell + n$ parton process.  (The $n$-jet
count excludes 
the remnant.)  We can get this matrix element easily from that for
$V\rightarrow \ell\ellbar + (n+1)$~partons, where $V$ represents a vector boson.
At next-to-leading 
order, the calculation of the same process also requires 
the tree-level matrix element for $V\rightarrow \ell\ellbar + (n+2)$~partons,
and the one-loop matrix element for $V\rightarrow \ell\ellbar + (n+1)$~partons.
(It furthermore requires a general formalism such as that of 
refs.~\cite{GG,GGK,FKS,CS} for 
cancelling the infrared singularities analytically while allowing a
numerical calculation of fully-differential observables.)

It is the calculation of the one-loop matrix elements that requires
the bulk of the theoretical effort in producing predictions for a new 
process.  In the traditional Feynman-diagram approach, even amplitudes
with four external partons\cite{ES} require hard, lengthy calculations;
and the situation only gets worse as the number of external legs grows.
The difficulties arise from the enormous number of diagrams, the large
amount of vertex algebra in each diagram, and the complexity of loop
integrals with many powers of the loop momentum in the numerator.  A
brute-force approach might easily lead to expressions thousands of times
larger than an appropriate representation of a result.

\def\qbar{{\overline q}}
The vastly more efficient methods developed in recent years start 
by taking advantage of earlier developments in both tree-level
calculations\cite{MP} and string-based methods for loop calculations~\cite{String}.
These include (a) color decomposition\cite{LoopColor}, (b) the spinor helicity 
method~\cite{SpinorHelicity},
(c) use of supersymmetric decompositions, (d) 
decomposition into `primitive' amplitudes, and the
use of permutation identities~\cite{QQGGG} 
that express subleading-color
amplitudes as a sum of permutations over `primitive'
amplitudes.  Primitive amplitudes correspond to color-ordered
amplitudes with a definite orientation of internal fermion
lines.  These fundamental building blocks may be computed efficiently 
using the twin tools
of unitarity-based sewing and factorization, developed in refs.~\cite{Unitarity}
and reviewed at length in ref.~\cite{AnnReview}.  These techniques have been
used extensively in a series of calculations, including the one-loop matrix 
elements~\cite{QQGGG}
for $0\rightarrow q\qbar ggg$, the all-multiplicity maximally-helicity violating
amplitudes in $N=4$ and $N=1$ supersymmetric theories~\cite{Unitarity}, and 
the one-loop matrix
elements for $\ell\bar{\ell}\rightarrow q\qbar q'\qbar'$~\cite{Zqqqq}
 and $\ell\bar{\ell}\rightarrow q\qbar gg$~\cite{Zqqgg}.

\section*{Calculation of Primitive Amplitudes}

A one-loop amplitude, in general, contains absorptive pieces.  
The corresponding dispersive terms
can be determined via dispersion relations from their cuts,
which are just given by products of tree amplitudes.  (The approach can
be extended beyond one loop; for first steps in this direction see
another contribution to this session~\cite{TwoLoop}.)  
In practice, one does not need to use dispersion relations
explicitly, but only implicitly, to determine the integral functions
that appear in any given amplitude, along with their coefficients.
The restricted number of integral functions that can appear implies 
that some cut-free pieces are also determined by this approach of
sewing tree amplitudes.  Indeed, in supersymmetric theories, the entire
amplitude is given by this technique.

The key point in the unitarity-based method is that we sew
tree {\it amplitudes\/}, not tree diagrams.  A calculation thus takes
advantage of all the cancellations and reductions in numbers of
terms that have already occurred in
the process of computing the tree amplitudes.

\def\oneloop{{\rm 1\hskip -1pt-\hskip -1ptloop}}
\def\tree{{\rm tree}}
A calculation using the unitarity-based method proceeds as follows.
We want to compute the coefficients $c_i$ of each of the box, triangle
or bubble integrals that might appear in the amplitude.  (The set of
possible integrals can
be determined by a `gedanken' reduction using Passarino-Veltman~\cite{PV}
and van~Neerven-Vermaseren~\cite{vNV}, or equivalent~\cite{Integrals} techniques.)
We consider in turn cuts in all possible
channels.  For a given channel, we form the cut in that channel, summing
over all intermediate states; this gives rise to a phase-space
integral of the form
\begin{equation}
\hskip -15mm\int d^D{\rm LIPS}(-\ell_1,\ell_2)\,
A^\tree_L(-\ell_1,\ldots,\ell_2) A^\tree_R(-\ell_2,\ldots,\ell_1)\,,
\label{cutEqn}
\end{equation}
where $\ell_{1,2}$ are the four-dimensional 
on-shell momenta crossing the cut, and
where the $A^\tree_{L,R}$ are the color-ordered tree subamplitudes 
on the two
sides of the cut. Using the Cutkosky rules, we can rewrite this expression
as the absorptive or imaginary part of a loop amplitude,
\begin{equation}
\left[ \int {d^D\ell_1\over (2\pi)^D} A^\tree_L {1\over \ell_2^2+i\varepsilon}
     A^\tree_R {1\over \ell_1^2+i\varepsilon}\right]_{\rm cut}\,.
\end{equation}
 In this expression, we may use the on-shell conditions $\ell_1^2 = 0 = \ell_2^2$
freely in all factors except the inserted propagators.
Using a power-counting theorem~\cite{Unitarity,AnnReview} in those cases where it applies
(or simply up to a polynomial ambiguity to be fixed as described below, in cases where
it does not apply), we can recover the real parts by dropping the
subscript ``cut''.  We then perform the usual reductions on the
resulting loop integral, and extract the coefficient of any integral function
containing a cut in the given channel.  (Functions which don't
contain a cut in the given channel should be dropped.)  Finally,
we reassemble the final answer by considering all channels.

The box integrals have cuts in more
than one channel.  Considering both channels provides us with a 
cross check --- the coefficients as computed in both channels must
agree --- or alternatively we could reduce the amount of work we must
do by considering only one channel.  The latter choice is particularly
appropriate when computing amplitudes in an $N=4$ supersymmetric gauge theory;
in this theory, 
all amplitudes can be written in terms of scalar
boxes. (We assume the use of a 
supersymmetrically-consistent version of dimensional regularization, such
as dimensional reduction, throughout.)

\def\e{\epsilon}
In non-supersymmetric theories, there are remaining cut-free or polynomial
terms which are not determined by sewing tree amplitudes in $D=4$.
It is possible to determine them if we use the cuts at ${\cal O}(\e)$ rather
than just to ${\cal O}(\e^0)$.  
The reason is that amplitudes in
a massless gauge theory have an over-all power of $(-s)^{-\e}$, where
$s$ is one of the momentum invariants of the external legs.  
The polynomial terms therefore do contain cuts
at ${\cal O}(\e)$, and can be deduced by sewing tree amplitudes, where
the momenta crossing the cuts are taken to be on-shell in $(4-2\e)$ dimensions
rather than four dimensions.  For scalars (and fermions~\cite{Fermions}), 
this is equivalent to computing
massive rather than massless amplitudes, followed by an appropriately
weighted integration over
the ``mass'' (really the $(-2\e)$-dimensional component of the momentum).

\def\Split{{C}}
However, for practical purposes, it is better to use collinear factorization 
to determine the polynomial terms.  This technique relies on the 
universal factorization of one-loop amplitudes in the collinear 
limit~\cite{Collinear,Unitarity},
which for leading-color amplitudes reads,
\begin{eqnarray}
&&
A^\oneloop_{n;1}(\cdots,a,b) \mathop{\longrightarrow}^{a\parallel b}
\nonumber\\
&& \hskip 5mm
 \sum_{\lambda_P=\pm} \Bigl[
  \Split^\tree_{-\lambda_P}(a^{\lambda_a},b^{\lambda_b})
      A^\oneloop_{n-1;1}(\cdots;P^{\lambda_P})
  +\Split^\oneloop_{-\lambda_P}(a^{\lambda_a},b^{\lambda_b})
      A^\tree_{n-1}(\cdots;P^{\lambda_P})\Bigr]
\end{eqnarray}
where the splitting amplitudes $\Split$ are universal functions depending
only on the collinear momenta and their helicities $\lambda$, 
and on the momentum fraction $z$ ($k_P = k_a+k_b$, $k_a = z k_P$).
In general, the terms deduced by sewing trees will not yield the correct
limit; one must add polynomial pieces.  One caveat in applying this
technique is that we have no proof that this determines the missing
polynomial pieces uniquely.  (It is
likely true for more than five external legs; for five external legs,
there is an ambiguity arising from the existence of a term which contains
no cuts but is collinear-finite.)  The results obtained using this technique
thus need to be checked
(for example, numerically) against results from another method.

\end{document}